# Neural Network-Aided BCJR Algorithm for Joint Symbol Detection and Channel Decoding

Wen-Chiao Tsai[1*], Chieh-Fang Teng[2*], Han-Mo Ou[1], An-Yeu (Andy) Wu[2], *Fellow*, *IEEE*
[1] Department of Electrical Engineering, National Taiwan University, Taipei, Taiwan
[2] Graduate Institute of Electrical Engineering, National Taiwan University, Taipei, Taiwan
{daniel, jeff}@access.ee.ntu.edu.tw, {b05901092, andywu}@ntu.edu.tw

*Abstract*—Recently, deep learning-assisted communication systems have achieved many eye-catching results and attracted more and more researchers in this emerging field. Instead of completely replacing the functional blocks of communication systems with neural networks, a hybrid manner of BCJRNet symbol detection is proposed to combine the advantages of the BCJR algorithm and neural networks. However, its separate block design not only degrades the system performance but also results in additional hardware complexity. In this work, we propose a BCJR receiver for joint symbol detection and channel decoding. It can simultaneously utilize the trellis diagram and channel state information for a more accurate calculation of branch probability and thus achieve global optimum with 2.3 dB gain over separate block design. Furthermore, a dedicated neural network model is proposed to replace the channel-model-based computation of the BCJR receiver, which can avoid the requirements of perfect CSI and is more robust under CSI uncertainty with 1.0 dB gain.

*Keywords—BCJR algorithm, Neural network, Symbol detection, Channel decoding, Turbo codes*

## I. Introduction

Due to its revolutionized breakthroughs in many fields, deep learning (DL) is considered as one of the most promising solutions to address the rigorous challenges toward 5G and beyond 5G (B5G) [1]-[2]. For example, convolutional neural networks are adopted as a powerful modulation classifier in [3]-[5]. For channel decoding, the authors in [6]-[8] proposed neural network-based decoders for BCH codes, polar codes, and turbo codes, which achieve better performance and faster convergence. Furthermore, for the dealing of channel interference, neural network-based equalizer [9] and symbol detection [10] are proposed with improved system performance and being more robust under inaccurate channel estimation, respectively.

Though DL is powerful and potential, it is inelegant to completely replace the well-designed functional blocks of communication systems with neural networks. It not only results in degraded system performance but also causes intolerable computational complexity. For example, poor generalization for unseen codewords and exponentially increased training complexity are met in [11]-[12] for the decoding of the codewords with longer block length. On the other hand, although the adopted recurrent neural network in [9] can jointly eliminate channel fading and utilize coding gain in advance, it is hard for such a huge model to do channel adaptation under dynamic environments. Thus, instead of completely replacing the functional blocks with neural networks, hybrid manner is a wiser approach to combine the advantages of both well-designed algorithms and neural networks, which can complement the defects of each other [10].

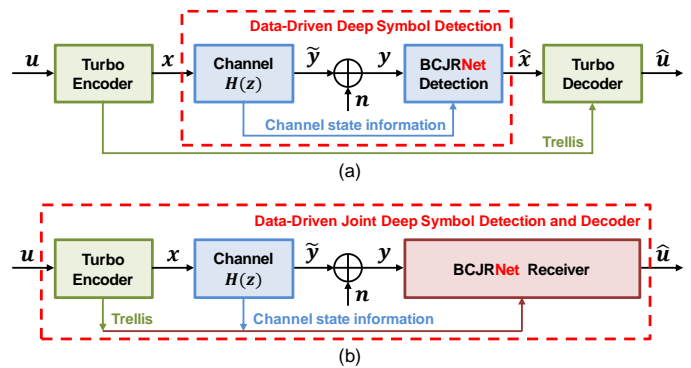

Fig. 1. Overview of (a) BCJRNet symbol detection [10], and (b) the proposed BCJRNet receiver with joint symbol detection and channel decoding.

In [10], a data-driven deep symbol detection based on the BCJR algorithm is proposed as shown in Fig. 1(a). Conventionally, the BCJR algorithm demands perfect channel state information (CSI) for the precise calculation of branch probability, which is not realistic and limits its application. Consequently, the authors in [10] remove BCJR's channel-model-dependence by replacing the part of CSI-based computations with a dedicated neural network model. Instead of using DL to replace the whole functional block, it achieves the performance guarantees and controllable model complexity with smaller model size. Thus, it also has better adaptability to channel variation. However, two issues should be addressed:

1) *Separate block design*: From Fig. 1(a), the estimated CSI and the code structure of the trellis diagram are independently utilized in the block of symbol detection and channel decoding, respectively, which results in a local optimum.

2) *Twice the hardware complexity*: Furthermore, for the BCJR symbol detection, it has comparable hardware complexity as the following BCJR decoder. Thus, the overall hardware complexity is doubled.

In this paper, keeping the design concept of hybrid fashion, we propose a data-driven BCJRNet receiver with joint symbol detection and channel decoding as shown in Fig. 1(b). Our main contributions are summarized as follows:

1) *Joint symbol detection and decoding*: From Fig. 1(b), the branch probability is jointly determined by the code structure and CSI, which is more robust under the noise environment and achieves a global optimum with 2.3 dB

[*]These two authors contributed equally.

This research work is financially supported by the MediaTek Inc., Hsinchu, Taiwan, under Grants MTKC-2019-0070. The second author is also sponsored by MediaTek Ph.D. Fellowship program.

gain over separate block design. Furthermore, the operations can be realized in one functional block, which reduces the hardware complexity.

2) *Data-driven BCJRNet receiver*: We further propose a dedicated neural network model to replace the channel-model-based computation of the BCJR receiver, which can avoid the requirements of perfect CSI and is more robust under CSI uncertainty with 1.0 dB gain.

The remainder of this paper is organized as follows: Section II briefly introduces the system model and the BCJR algorithm. Section III presents the proposed BCJR receiver with its data-driven extension. Section IV shows the experimental results and analysis. Finally, we conclude this paper in Section V.

## II. BACKGROUND

### A. System Model

The system model is depicted in Fig. 1. At the transmitter side, the $K$ information messages $u$ are first encoded as $x$ by the turbo encoder that contains two identical recursive systematic convolutional (RSC) encoders as shown in Fig. 2(a). The first encoder generates two sequences of systematic bits $x^s = u$ and parity bits $x^{p_1}$, and the second encoder generates another sequence of parity bits $x^{p_2}$ based on the input of the interleaved message bits $u' = \prod(u)$ [13]. Then, the BPSK modulated codeword $x = [x_0, x_1, x_2, ..., x_{3K-1}] = [x_0^s, x_0^{p_1}, x_0^{p_2}, ..., x_{K-1}^s, x_{K-1}^{p_1}, x_{K-1}^{p_2}]$ consisting of $N = 3K$ bits is sent to the channel. Throughout this paper, channel fading is considered. Inter-symbol interference (ISI) and additive white Gaussian noise (AWGN) jointly contribute to channel distortion and the received signal can be expressed as:

$$y_k = \sum_{i=0}^{L-1} x_{k-i} \times h_i + w_k, k = 0, ..., 3K - 1, \quad (1)$$

where $L$ is the length of the response, $h_k \triangleq e^{-\gamma k}$ is the channel impulse response with exponentially decaying profile and $w_k \sim N(0, \sigma^2)$ denotes i.i.d. additive white Gaussian noise.

At the receiver side, the BCJR symbol detection is firstly applied to recover the distorted signal. Then, based on the reconstructed signal $\hat{x}$, a turbo decoder estimates the log-likelihood ratios (LLRs) for the original information messages $u$ by iteratively adopting the BCJR algorithm as shown in Fig. 2(b) and Fig. 2(c) [14]. The soft output of LLRs indicates the probability of the corresponding bits being 0 or 1 and thus can be utilized to determine the estimated message bits $\hat{u}$.

### B. BCJR Algorithm [14]

The BCJR algorithm is a maximum *a posteriori* probability (MAP) algorithm and can be applied to both the tasks of symbol detection and decoding, which are defined on factor graphs and trellises. For convenience, we will assume the mapping $\{0 \leftrightarrow -1, 1 \leftrightarrow +1\}$ for message bits $u$ and codeword $x$. Take turbo decoder as an example, the BCJR algorithm iteratively computes the log *a posteriori* probability ratio of message bits as follows:

$$L(u_k) = \ln \frac{P(u_k = +1|y)}{P(u_k = -1|y)} = \ln \frac{\sum_{S^+} P(s', s, y)}{\sum_{S^-} P(s', s, y)}, \quad (2)$$

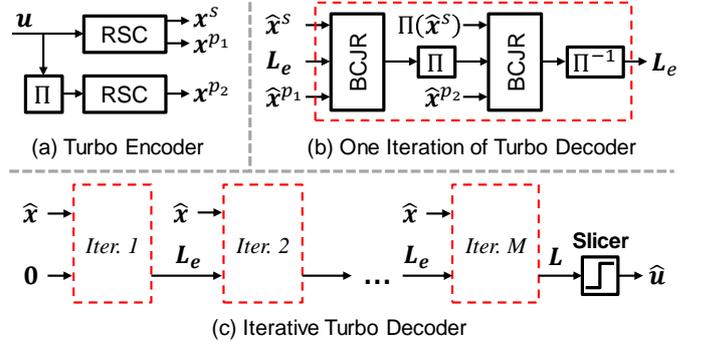

Fig. 2. (a) Turbo encoder with two recursive systematic convolutional (RSC) encoders, (b) operations in one iteration of turbo decoder, and (c) iterative turbo decoder with $M$ iterations.

where $s'$ and $s$ represent the states of the RSC encoder at time $k - 1$ and $k$, respectively, and $y$ for decoder 1 (decoder 2) is the sequence made up of the reconstructed systematic bits $\hat{x}^s$ ($\prod(\hat{x}^s)$) and its corresponding parity bits $\hat{x}^{p_1}$ ($\hat{x}^{p_2}$). $S^+$ and $S^-$ are the set of ordered pairs $(s', s)$ corresponding to all states transit from $s'$ to $s$ caused by $u_k = +1$ and $u_k = -1$, respectively. The joint probability $P(s', s, y)$ can be further decomposed by Bayes's rule:

$$P(s', s, y) = \alpha_{k-1}(s')\gamma_k(s', s)\beta_k(s), \quad (3)$$

where $\alpha_{k-1}(s') = P(s', y_{<k})$, $\gamma_k(s', s) = P(y_k, s|s')$, and $\beta_k(s) = P(y_{>k}|s)$ represent forward recursion, branch probability, and backward recursion, respectively, and can be calculated as follows:

$$\alpha_k(s) = \sum_{s'} \alpha_{k-1}(s')\gamma_k(s', s), \alpha_0(s) = \begin{cases} 1, s = 0 \\ 0, s \neq 0 \end{cases}, \quad (4)$$

$$\gamma_k(s', s) = \exp\left\{-\frac{\|\hat{x}_k - x_k\|^2}{2\sigma^2} + \frac{u_k L_e(u_k)}{2}\right\}, \quad (5)$$

$$\beta_{k-1}(s') = \sum_s \beta_k(s)\gamma_k(s', s), \beta_N(s) = \begin{cases} 1, s = 0 \\ 0, s \neq 0 \end{cases}, \quad (6)$$

where $\hat{x}_k \triangleq [\hat{x}_k^s \; \hat{x}_k^{p_1}]$ and $x_k \triangleq [x_k^s \; x_k^{p_1}]$ represents the reconstructed codeword and the transmitted codeword, respectively. $L_e(u_k)$ is the extrinsic information received from the companion decoder. As shown in Fig. 2(b) and Fig. 2(c), the two BCJR decoders iteratively transmit the interleaved (deinterleaved) extrinsic information $L_e(u_k)$ to each other and result in a more accurate estimation of $L(u_k)$.

To reduce computational complexity, the operations of multiplication can be substituted by addition as introduced in [15]. Therefore, $\alpha_k(s), \gamma_k(s', s)$, and $\beta_{k-1}(s')$ are calculated in logarithmic domain with the Jacobian logarithmic function as follows:

$$A_k(s) = \ln \alpha_k(s) = \max_{s'}^*\{A_{k-1}(s') + \Gamma_k(s', s)\}, \quad (7)$$

$$\Gamma_k(s', s) = \ln \gamma_k(s', s) = -\frac{\|\hat{x}_k - x_k\|^2}{2\sigma^2} + \frac{u_k L_e(u_k)}{2}, \quad (8)$$

$$B_{k-1}(s') = \ln \beta_{k-1}(s') = \max_s^*\{B_k(s) + \Gamma_k(s', s)\}, \quad (9)$$

where $\max^*(a, b) = \max(a, b) + \ln(1 + e^{-|a-b|})$ or can be further simplified as $\max(a, b)$, which are log-MAP algorithm

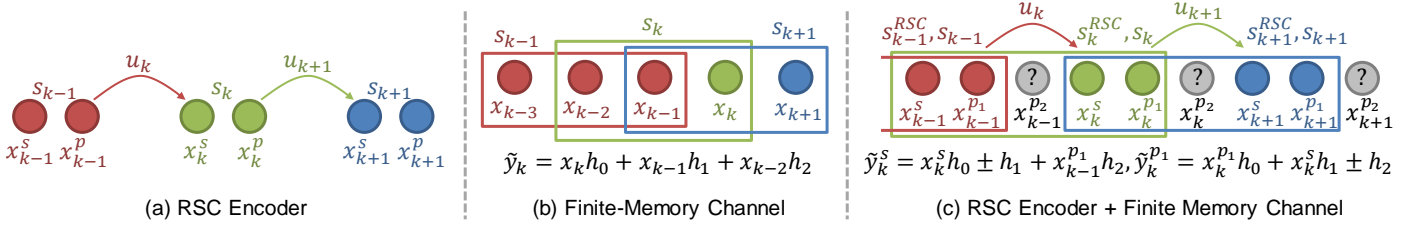
Fig. 3. State transitions of different schemes: (a) RSC encoder, (b) finite-memory channel with $\mathbf{h} = [h_0, h_1, h_2]$, and (c) the proposed joint RSC encoder and finite-memory channel for the case of first decoder.

or max-log-MAP algorithm, respectively. For the following experiments in this work, we adopt log-MAP algorithm.

### C. BCJRNet Symbol Detection [10]

The BCJR algorithm can also be applied to the problem of symbol detection under a finite-memory channel model. For turbo decoding, the state is defined based on the code structure of the trellis diagram as shown in Fig. 3(a) and Table. I(a). However, for the task of symbol detection, the state is defined as $s_k \triangleq (x_k, \ldots, x_{k-L+1})$, where $L$ is the channel length as shown in Fig. 3(b) and Table. I(b). Then, symbol detection can be realized by (2) with the newly defined state. However, in order to calculate the joint probability in (3), prior knowledge of the channel model and its parameters, such as full CSI, is required; otherwise, the branch probability $\gamma_k(s', s)$ cannot be realized. However, in some cases, such accurate prior knowledge may not be available or is costly to acquire, which degrades the system performance.

To address this challenge, the authors in [10] proposed a dedicated DNN model to learn the channel-model-based computations $\gamma_k(s', s)$ in a data-driven fashion and the remaining flow of the BCJR algorithm, such as forward recursion and backward recursion, is unchanged. By doing so, the DNN model can easily learn the CSI from a small dataset and predict branch probability without requiring an accurate channel model and corresponding parameters. For more details, please refer to [10].

## III. PROPOSED BCJRNET RECEIVER WITH JOINT SYMBOL DETECTION AND CHANNEL DECODING

### A. BCJR Receiver with Joint Symbol Detection and Channel Decoding

In communication systems, in addition to the block of symbol detection, channel coding is widely adopted to correct the error bits and improve the system performance as shown in Fig. 1(a). However, according to [9], though conventional communication systems with separated block designs are more controllable, the coding gain can only be exploited in the stage of channel decoding, which results in local optimum and sacrifices the overall system performance. Furthermore, in the light of the BCJR algorithm can be applied to both symbol detection and channel decoding, we propose a BCJR receiver for joint symbol detection and channel decoding by simultaneously considering the trellis diagram and channel state information.

As in the Turbo decoder, the BCJR receiver is composed of two BCJR decoders. Each constituent decoder receives extrinsic information from its companion decoder along with the channel state information. The two decoders then iteratively estimate the LLR $L(u_k)$ given the received codeword $\mathbf{y}$. In order to calculate the *a posteriori* probability $P(u_k|\mathbf{y})$, we need to redefine the state for the trellis in the BCJR decoder since the successive bits of received codeword $\mathbf{y}$ is not only correlated by the trellis diagram of encoder but also associated with the channel fading.

**TABLE I**. State transitions of different schemes.

| (a) RSC Encoder with Generator Matrix $[1, 1+D^2/1+D+D^2]$ | | | | | |
|---|---|---|---|---|---|
| $s_{k-1}$ | | 00 | 01 | 10 | 11 |
| $s_k$ | $u_k = 0$ | 00 | 10 | 11 | 01 |
| | $u_k = 1$ | 10 | 00 | 01 | 11 |
| $x_k^s x_k^p$ | $u_k = 0$ | 00 | 00 | 01 | 01 |
| | $u_k = 1$ | 11 | 11 | 10 | 10 |

| (b) Finite-Memory Channel with $\mathbf{h} = [h_0, h_1, h_2]$ | | | | | | | | | |
|---|---|---|---|---|---|---|---|---|---|
| $s_{k-1}$ | | 000 | 001 | 010 | 011 | 100 | 101 | 110 | 111 |
| $s_k$ | $x_k = -1$ | 000 | 010 | 100 | 110 | 000 | 010 | 100 | 110 |
| | $x_k = 1$ | 001 | 011 | 101 | 111 | 001 | 011 | 101 | 111 |
| $\tilde{y}_k$ | | $x_k h_0 + x_{k-1} h_1 + x_{k-2} h_2$ | | | | | | | |

| (c) RSC Encoder + Finite-Memory Channel | | | | | | | | | |
|---|---|---|---|---|---|---|---|---|---|
| $s_{k-1}$ | | 000 | 001 | 010 | 011 | 100 | 101 | 110 | 111 |
| $s_k$ | $u_k = 0$ | 000 | 100 | 100 | 000 | 110 | 010 | 010 | 110 |
| | $u_k = 1$ | 001 | 101 | 101 | 001 | 111 | 011 | 011 | 111 |
| $\tilde{y}_k^s, \tilde{y}_k^{p_1}$ | | $x_k^s h_0 \pm h_1 + x_{k-1}^{p_1} h_2, x_k^{p_1} h_0 + x_k^s h_1 \pm h_2$ | | | | | | | |

For convenience, we consider the BCJR receiver applied to a rate-1/3 turbo encoder over a finite-memory channel with memory length $L = 3$. Generalizing to other code rates or memory lengths is straightforward. Similar to the transmitted codeword $\mathbf{x}$ mentioned in Subsection II.A, the received codeword $\mathbf{y}$ is defined as $\mathbf{y} = [y_0, y_1, y_2, \ldots, y_{3K-1}] = [y_0^s, y_0^{p_1}, y_0^{p_2}, \ldots, y_{K-1}^s, y_{K-1}^{p_1}, y_{K-1}^{p_2}]$, where $y_k$ can be obtained from (1) by substituting $L = 3$ as follows:

$$y_k = x_k h_0 + x_{k-1} h_1 + x_{k-2} h_2 + w_k. \quad (10)$$

We denote the input to the first decoder and the second decoder as $\mathbf{y}_1 = [\mathbf{y}_1^1, \mathbf{y}_2^1, \ldots, \mathbf{y}_K^1]$ and $\mathbf{y}_2 = [\mathbf{y}_1^2, \mathbf{y}_2^2, \ldots, \mathbf{y}_K^2]$, respectively, where $\mathbf{y}_k^1 \triangleq [y_k^s, y_k^{p_1}]$, $\mathbf{y}_k^2 \triangleq [y_k^{s'}, y_k^{p_2}]$ and $\mathbf{y}^{s'} = \prod(\mathbf{y}^s)$.

In the following discussion, we will show how to define the states in the BCJR decoder and how to calculate the branch probability based on the newly defined states. We will take the first decoder as an example, and the second decoder can be derived similarly. The relation between the $k$th input $\mathbf{y}_k^1$ and the transmitted codeword $\mathbf{x}$ can be derived from (10) as follows:

$$y_k^s = x_k^s h_0 + x_{k-1}^{p_2} h_1 + x_{k-1}^{p_1} h_2 + w_{3k}, \quad (11)$$

$$y_k^{p_1} = x_k^{p_1} h_0 + x_k^s h_1 + x_{k-1}^{p_2} h_2 + w_{3k+1}, \quad (12)$$

which is depicted in Fig. 3(a). Notice that if we define the state to be the contents of the memory elements in the RSC encoder as in the case of Turbo decoder, the joint probability $P(s', s, \mathbf{y})$ cannot be factorized to the form in (3) since the received codeword at time $k$ ($\mathbf{y}_k^1$) now depends on both the transmitted codewords at time $k$ ($x_k^s, x_k^{p_1}$) and $k-1$ ($x_{k-1}^{p_1}$) as a direct consequence of the ISI channel.

To tackle this problem, we need to extend our state by combining the successive two states to form a bigger one. Therefore, it contains the information of the transmitted codeword and then we can utilize CSI for the calculation of the branch probability. We denote the state of the RSC encoder at time $k$ as $s_k^{RSC}$. Then, the newly defined state at time $k$ is $s_k \triangleq (s_{k-1}^{RSC}, u_k)$ since the transmitted codeword at time $k$ is the output of the RSC encoder with the current state $s_{k-1}^{RSC}$ and input $u_k$. Besides, the transmitted codeword at time $k-1$ can also be indicated by the newly defined state at time $k-1$.

After defining the state for the BCJR algorithm, we need to determine the calculation of the branch probability $\gamma_k(s', s)$ used for forward recursion and backward recursion. We define the noiseless version of $\mathbf{y}_k^1$ to be $\widetilde{\mathbf{y}}_k^1 \triangleq [\widetilde{y}_k^s, \widetilde{y}_k^{p_1}]$, where

$$\widetilde{y}_k^s = x_k^s h_0 + x_{k-1}^{p_2} h_1 + x_{k-1}^{p_1} h_2, \quad (13)$$

$$\widetilde{y}_k^{p_1} = x_k^{p_1} h_0 + x_k^s h_1 + x_{k-1}^{p_2} h_2. \quad (14)$$

For the transmitted codeword $x_{k-1}^{p_2}$ in (13) and (14), since it is generated by the second encoder and will be only considered in the second decoder, we assume it takes the values in $X = \{\pm 1\}$ with equal probabilities. Then $\gamma_k(s', s)$ can be calculated from (5) as follows:

$$\gamma_k(s', s) = \frac{1}{2} \sum_{x_{k-1}^{p_2} \in X} \exp\left[-\frac{\|\mathbf{y}_k^1 - \widetilde{\mathbf{y}}_k^1\|^2}{2\sigma^2} + \frac{u_k L_e(u_k)}{2}\right]. \quad (15)$$

By substituting (15) into (8) and ignoring the term independent of $u_k$, we can get the logarithmic domain branch metric as below:

$$\Gamma_k(s', s) = \max_{x_{k-1}^{p_2} \in X}^* \left(-\frac{\|\mathbf{y}_k^1 - \widetilde{\mathbf{y}}_k^1\|^2}{2\sigma^2}\right) + \frac{u_k L_e(u_k)}{2}, \quad (16)$$

where the function $\max^*(\cdot)$ is used to further simplify the equation. For clear description, we compare the state transition of different schemes as visualized in Fig. 3 and summarized in Table. I. The transmitted codeword $x_{k-1}^{p_2}$ in (13) and (14) is replaced with $\pm 1$ to indicate the possible values it may take. With the newly defined state and branch probability, we apply the BCJR algorithm for the forward and backward recursions and our BCJR receiver can estimate the LLR of the information bits via iterative decoding process of the two BCJR decoders.

### B. Neural Network-Aided BCJR Receiver

For our BCJR receiver to function properly, we have to provide accurate channel state information (CSI) for the calculation of branch probabilities in the BCJR decoders according to (13) and (14). However, in real applications, accurate prior information of the channel may not be available, and a bad estimation of CSI will directly degrade the performance of the BCJR receiver.

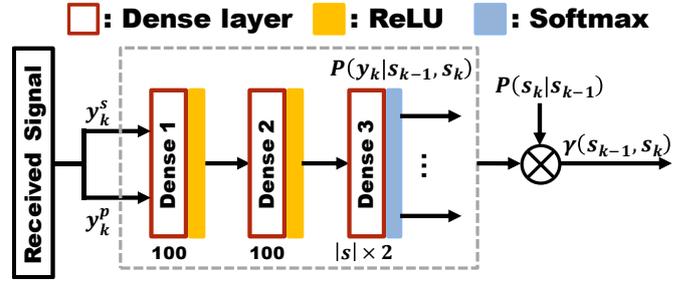

Fig. 4. The calculation of branch probability for the proposed BCJRNet receiver.

In [10], BCJRNet symbol detection is proposed to successfully combine the advantages of the BCJR algorithm and neural network to avoid the requirement of full CSI. This hybrid fashion inspires us to address the challenge of unavailable CSI by taking advantage of neural networks. Consequently, the calculation of the branch probability can be realized in a data-driven manner, and then the acquisition of branch probability in both the forward and backward recursions can be carried out by inferencing the well-trained model.

In order to calculate the branch probability with a neural network model, $\gamma_k(s', s)$ cab be rewritten as below:

$$\gamma_k(s', s) = P(\mathbf{y}_k|s', s) P(s|s'), \quad (17)$$

where $P(s|s')$ is the probability of the event $s' \to s$. Note that $P(s|s') = 0$ if $s$ is not a valid state from $s'$. Otherwise, $P(s|s')$ can be obtained from the extrinsic information received from the companion decoder in the iterative decoding process. Since $\mathbf{y}_k$, the received codeword at time $k$, takes continuous values and the number of events caused by the transition to a valid state $s$ from $s'$ is finite, it is straightforward to learn a model which takes $\mathbf{y}_k$ as input and outputs $P(\mathbf{y}_k|s', s)$ for each valid transition.

Consequently, instead of training an end-to-end model that imitates the behaviors of the BCJR receiver, we only replace the part of the channel-model-based computation of $P(\mathbf{y}_k|s', s)$ with neural networks for the computation of branch probability. Therefore, we can train a relatively compact model with a small set of training data. The calculation of branch probability for the proposed BCJRNet receiver is shown in Fig. 4. The dedicated model has three fully-connected layers and the values below the dense layer represent the number of nodes. The nonlinear function used by the first two layers, Rectified Linear Units (ReLUs), is defined as:

$$f_{\text{ReLU}}(x) = \max\{0, x\}, \quad (18)$$

which can result in a nonlinear system and augment the learning capability of the neural network. The softmax activation is used at the output layer to normalize the estimated branch probability for each state and can be given by:

$$f_{\text{Softmax}}(\mathbf{x}_i) = \frac{e^{x_i}}{\sum_{j=0}^{2 \times |s|-1} e^{x_j}}. \quad (19)$$

**TABLE II.** Simulation parameters.

| Modulation category | BPSK |
|---|---|
| Message length ($k$) | 100 |
| RSC encoder | $[1, 1+D^2/1+D+D^2]$ |
| Interleaver | Random |
| BCJR iterations | 6 |
| Fading channel | $[1, e^{-1}, e^{-2}]$ |
| CSI uncertainty ($\sigma_e^2$) | 0.1 |
| Testing SNR | -6, -4, -2, 0, 2, 4, 6 (dB) |
| Training SNR | 0 (dB) |
| Training codeword | 10,000 |
| Optimizer | SGD with Adam |
| Training and testing environment | Deep learning library of PyTorch with NVIDIA RTX 8000 GPU |

The training data is generated based on the decoding process of the BCJR receiver, where the input data is the received codeword at time $k$ and the corresponding labeled data is the normalized calculation of $P(\mathbf{y}_k|s',s)$ as the following equation:

$$P(\mathbf{y}_k|s',s) = \frac{1}{2}\sum_{x_{k-1}^{p_2}\in X}\exp\left[-\frac{\|\mathbf{y}_k^1-\tilde{\mathbf{y}}_k^1\|^2}{2\sigma^2}\right]. \quad (20)$$

For the training process, we utilize the Kullback-Leibler divergence (KLD) loss to train our model since KLD is a useful distance metric to measure the difference between two probability distributions. The KLD loss between the labeled probability $\mathbf{p}$ and the predicted probability $\mathbf{q}$ is defined as:

$$KL(\mathbf{p}||\mathbf{q}) = \sum_{i=0}^{2\times|s|-1} p_i \log\frac{p_i}{q_i}. \quad (21)$$

## IV. EXPERIMENTAL RESULTS AND ANALYSIS

As mentioned in Section II.A, the channel considered in this paper is a fading channel with AWGN, where $\gamma$ is set to 1. The length of message bits $k$ is set to 100 and is encoded to 300 bits by turbo encoder. In our experiments, we focus on the comparison between the different schemes of receiver design. The transmitter is unchanged throughout the experiments as shown in Fig. 1. For each point, there are at least 400 error bits to make sure the results are stable enough. The simulation setup is summarized in Table II.

### A. Comparison of Proposed BCJR Receiver and Prior Work

The first experiment is to evaluate the performance between the proposed BCJR receiver and prior work [10], which has two separate blocks of symbol detection and channel decoding as shown in Fig. 1(a). Besides, the performance of DeepTurbo [16], applying 2-layer bidirectional recurrent neural network (RNN) to replace the block of BCJR algorithm, is also evaluated. We also examine the performance of our BCJRNet receiver. Note that for the training of the NN model, we set the SNR value to 0dB. However, during testing, the SNR ranges from -6dB to 6dB, which can evaluate the generalization of our model under different SNR values.

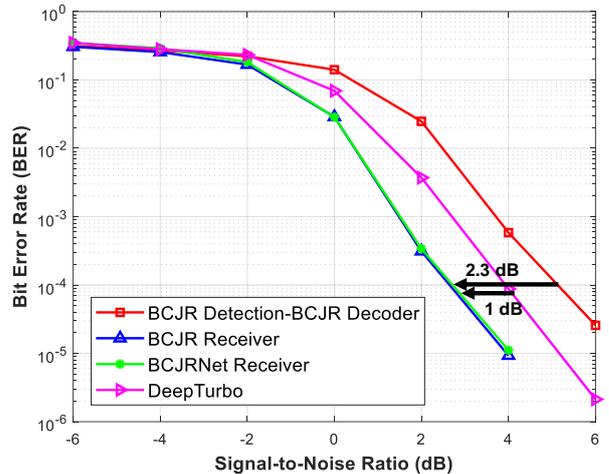

Fig. 5. Comparison of BER between the proposed BCJR receiver, prior work with separate symbol detection and channel decoding [10], and DeepTurbo [16].

From Fig. 5, we can observe that there is a significant performance gap between the proposed BCJR receiver and the separate design of symbol detection and decoding. It demonstrates that our proposed approach can jointly utilize the trellis diagram and CSI for a more accurate estimation of branch probability. Instead, the separate design deals with symbol detection and channel decoding independently, where the code structure of trellis cannot be utilized to protect the calculation of branch probability from channel noise.

Besides, we can observe that the BCJRNet receiver has almost the same performance as the BCJR receiver under the whole SNR ranges. Only an ignorable gap appears at high SNR ranges, which may stem from that the training data is collected under 0dB. However, the results still demonstrate that our model can generalize to a wide range of SNR values and have 2.3 dB gain over separate block design. Furthermore, our BCJRNet receiver also has 1 dB improvement compared to DeepTurbo [16]. In our internal experiments, the performance of DeepTurbo can approach our BCJRNet receiver as the number of training data increases significantly. It demonstrates that DeepTurbo can also learn joint symbol detection and channel decoding by using complicated RNN model with a lot of training data. On the other hand, the hybrid approach can achieve performance guarantees and practical model complexity by taking advantages of both well-designed algorithms and neural networks.

### B. Comparison of Proposed BCJR Receiver Under CSI Uncertainty

In most cases, perfect CSI is not available, which results in an inaccurate calculation of branch probability and thus degrades the system performance. In this experiment, to simulate CSI uncertainty, we refer to [10] by adding a zero-mean Gaussian noise with variance $\sigma_e^2$ to the perfect CSI of $\mathbf{h}$. Therefore, there is a mismatch between the actual channel impulse response and the CSI utilized for the calculation of branch probability. On the other hand, the proposed BCJRNet is trained with samples from different realizations of noisy $\mathbf{h}$.

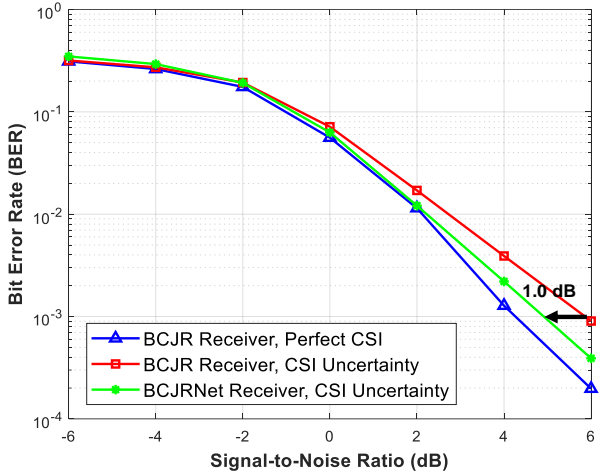

Fig. 6. Comparison of BER between BCJR receiver and BCJRNet receiver under channel state information (CSI) uncertainty.

From Fig. 6, we can observe that there is an obvious performance gap between the BCJR receiver with perfect CSI and under CSI uncertainty, which is caused by inaccurate calculation of branch probability. However, the BCJRNet receiver can still retain a similar performance as the BCJR receiver with perfect CSI. The results prove that the proposed BCJRNet receiver is more robust to CSI uncertainty with 1.0 dB gain, which also demonstrates the benefit of a hybrid approach to combine the experts of the BCJR algorithm and neural networks.

## V. CONCLUSIONS

In this work, we propose a BCJR receiver for joint symbol detection and channel decoding. By simultaneously considering the code structure of the trellis diagram and CSI, our approach can achieve 2.3 dB gain against separate block design. Furthermore, we propose a dedicated neural network model to replace the channel-model-based computation of the BCJR receiver, which can avoid the requirements of perfect CSI and is more robust under CSI uncertainty. Our approach demonstrates the advantages of end-to-end system design as well as the hybrid manner to combine the benefits of communication algorithms and neural networks.